\documentstyle[sprocl]{article}

\input{epsf}

\def\ra{\rightarrow}

\def\be{\begin{equation}}
\def\ee{\end{equation}}
\def\bea{\begin{eqnarray}}
\def\eea{\end{eqnarray}}
\begin{document}
\def\no{\nonumber}
\def\qu{\quad}
\def\qb{\bar{q}}
\def\qbm{\bar{\mbox{q}}}
\def\la{\langle}
\newcommand{\cor}[1]{\left<{#1}\right>}
\newcommand{\gm}{\gamma}
\renewcommand{\be}{\begin{eqnarray}}
\renewcommand{\ee}{\end{eqnarray}}
\renewcommand{\th}{\theta}
\newcommand{\Sg}{\Sigma}
\newcommand{\dl}{\delta}
\newcommand{\SSg}{\tilde{\Sigma}}
\newcommand{\eq}{\begin{equation}}
\newcommand{\eqx}{\end{equation}}
\newcommand{\eqn}{\begin{eqnarray}}
\newcommand{\eqnx}{\end{eqnarray}}
\newcommand{\ben}{\begin{eqnaray}}
\newcommand{\een}{\end{eqnarray}}
\newcommand{\f}[2]{\frac{#1}{#2}}
\renewcommand{\ra}{\longrightarrow}
\newcommand{\GG}{{\cal G}}
\renewcommand{\AA}{{\cal A}}
\newcommand{\GR}{G(z)}
 \newcommand{\MM}{{\cal M}}
\newcommand{\BB}{{\cal B}}
\newcommand{\ZZ}{{\cal Z}}
\newcommand{\DD}{{\cal D}}
\newcommand{\HH}{{\cal H}}
\newcommand{\RR}{{\cal R}}
\newcommand{\VV}{{\cal V}}
\newcommand{\GT}{{\cal G}_1 \otimes {\cal G}_2^T}
\newcommand{\GGb}{\bar{{\cal G}}^T}
\newcommand{\Du}{{\cal D}_1}
\newcommand{\Dl}{{\cal D}_2}
\newcommand{\zb}{\bar{z}}
\newcommand{\trqq}{\tr_{q\bar{q}}}
\newcommand{\arr}[4]{
\left(\begin{array}{cc}
#1&#2\\
#3&#4
\end{array}\right)
}
\newcommand{\arrd}[3]{
\left(\begin{array}{ccc}
#1&0&0\\
0&#2&0\\
0&0&#3
\end{array}\right)
}
\newcommand{\tr}{\mbox{\rm tr}\,}
\newcommand{\trn}{\mbox{\rm tr}_N\,}
\newcommand{\One}{\mbox{\bf 1}}
\newcommand{\pauli}{\sg_2}
\newcommand{\corr}[1]{\la{#1}\rangle}
\newcommand{\br}[1]{\overline{#1}}
\newcommand{\phib}{\br{\phi}}
\newcommand{\psib}{\br{\psi}}
\newcommand{\lm}{\lambda}
\newcommand{\ksi}{\xi}
\newcommand{\Gb}{\br{G}}
\newcommand{\Vb}{\br{V}}
\newcommand{\Gm}{G_{q\br{q}}}
\newcommand{\Vm}{V_{q\br{q}}}
\newcommand{\ggd}[2]{\GG_{#1}\otimes\GG^T_{#2}\Gamma}
\newcommand{\noi}{\noindent}

\newcommand{\refnote}[1]{#1}
	
\title{RANDOM MATRICES AND CHIRAL SYMMETRY in QCD
\protect\footnote{Talk presented by MAN  at "Hadrons in Dense Matter", 
Seoul, 1997.}}

\author{Romuald A. Janik}

\address{Department of Physics, Jagellonian University, 
  30-059 Krak\'{o}w, Poland}

\author{Maciej A. Nowak}

\address{Department of Physics, Jagellonian University, 
  30-059 Krak\'{o}w, Poland \\
GSI, Planckstr. 1, D-64291 Darmstadt, Germany \\
Institut f\"{u}r Kernphysik, TH Darmstadt, D-64289
Darmstadt, Germany}

\author{G\'{a}bor Papp}

\address{ITP, University of Heidelberg, Philosophenweg 19, D-69120 
Heidelberg, Germany \\
  Institute for Theoretical Physics, E\"{o}tv\"{o}s
  University Budapest, Hungary}

\author{Ismail Zahed}

\address{Department of Physics, SUNY, Stony Brook, NY 11794, USA}

%\thanks{Invited talk  by MAN  at ``Hadrons in Dense Matter''
%Seoul,  
%1997.}

\maketitle
%\abstract{%
%Abstarct goes here
%}
%\PACS{PACS go here}

\section{INTRODUCTION}

In this talk we review some recent results from random matrix models as
applied to some non-perturbative issues in QCD. All of the issues we will
discuss touched upon the important phenomenon related to the spontaneous 
breaking of chiral symmetry.
%The afore mentioned insights are:\\
%1. Spontaneous breakdown of chiral symmetry and disorder.\\ 
%2. Universal microscopic 
%  properties of the eigenvalues of the Dirac operator in the vacuum.\\
%3.  Universal microscopic 
%  properties of the eigenvalues of the Dirac operator in matter.\\
%4. Structural changes of the Dirac spectrum - finite temperature.\\
%5. Structural changes of the Dirac spectrum  - finite baryonic
%density  - ``phony vacua''\\
%6. Structural changes of the Dirac spectrum  - finite baryonic
%density  - ``true vacua'' .\\
%7. Phase diagram.\\ 
%8. Critical parameters.\\
%9. Critical exponents.\\
%10. $U(1)_A$ problem.\\
%11. Screening of the pseudoscalar susceptibility.\\
%12. Strong CP violation (finite $\theta$).\\   

Instead of using the standard lore of Green's functions
in random matrix models, we will instead choose to work
with their functional inverse or Blue's functions as
defined by~\cite{ZEE}. This way of doing things sheds
much insight into the physics of random matrix models,
and is probably the most ``natural'' and user-friendly 
way of introducing the formal, but powerful mathematical 
concept of free random variables~\cite{VOICULESCU}. 
In this talk we skip almost all  mathematical details
(referring the interested reader to the published work), 
and append some formal and mathematically relevant issues 
to an Appendix.

Exploring new and pertinent mathematical issues in the realm 
of frontier physics has been one of Mannque Rho specialty. We
hope that the present lecture which combines topics of current
physical interest in chiral symmetry breaking and QCD, with 
novel mathematical concepts, can be regarded as a small tribute
to Mannque Rho's 60th birthday. 

\section{Chiral Symmetry Breakdown and Random Matrix Models}
\subsection{Disorder}

The first important question is: why  random matrix models
have anything in common with  QCD? The answer will come from
the generic aspects of the way chiral symmetry is spontaneously
broken in the vacuum, and the fact that sigma-models are good
starting point. In the process, we will show that the results 
of RMM fall into universal (microscopic regime) and non-universal
(macroscopic regime) both of which give us much insights to the
intricacies of the QCD Dirac spectrum in a finite volume. We would
like to stress from the onset that the universal results provided
by RMM are not only applicable to QCD but to any effective model of 
QCD that breaks spontaneously chiral symmetry.

In the macroscopic regime qualitative but hopefully generic 
features of non-perturbative results can be probed with much
insights to the important parameters at work in reaching the
thermodynamical limit. QCD in a finite volume as probed by 
current lattice simulations is in many ways a complex and
disordered system. RMM are attractive tools to model disorder
in the presence of external and deterministic parameters such as
masses, temperature, chemical potential, and vacuum angle. 
Changes in these parameters introduce specific signatures on the
QCD Dirac spectrum, some of which may be generic enough to be 
modeled by pertinent matrices. As most of the physics retained
in RMM is soft, the crucial assumption relies on the decoupling of 
soft and hard modes in QCD. The many qualitative agreements between
RMM results and lattice simulations suggest that this may be the case.
A thorough analysis can be achieved by thorough lattice cross-checks 
between cooled and uncooled simulations.

Now, consider the Banks-Casher relation~\cite{BANKSCASHER}
\be
\corr{\bar{q}q}=-\frac{\nu(\lambda=0)}{\pi V_4}.
\label{BC}
\ee
 This remarkable relation links the order parameter of chiral symmetry
breakdown, i.e. quark condensate $\corr{\bar{q}q}$, to the average spectral
density $\nu$ 
of the eigenvalues  near $\lambda=0$ of the massless Dirac operator
$iD\!\!\!\!/(A)$. 
 The presence of the Euclidean volume $V_4$ in 
(\ref{BC}) is not accidental. It points out, that
the average spectral density of the Dirac operator at zero virtuality
has to grow like volume $V_4$, in order to provide the mechanism for
non-zero condensate in the thermodynamical limit. Free quarks in 
a finite volume $V$ exhibit at zero virtuality a density $\sqrt[4]{V_4}$
which is not enough in the thermodynamical limit to cause the spontaneous 
breaking of chiral symmetry.

We note the similarity of Banks-Casher relation to Kubo-Greenwood 
formula for direct current conductivity $\sigma$
\be
\sigma=D \rho(E_F)   
\ee
relating the conductivity to density of electronic 
states at the Fermi level, via the diffusion constant D. 
This analogy suggests, that the regime of spontaneous 
breakdown of  chiral symmetry is actually diffusive and 
originates from delocalization of quark eigenmodes. A
thorough characterization of this regime and others may be found
in our recent work~\cite{USDISORDER}.

In real QCD, the soft modes are likely to set in around instantons
configurations or any of their topological relatives, e.g. monopoles.
Cooling of the lattice has been shown to reproduce  the gross features
of QCD in the infrared limit, so a simplification of the quark modes
to the ones near the zero virtuality regime may be qualitatively legitimate
for bulk observables and Dirac spectra. This is minimally achieved through the 
use of RMM. The latters capture the essentials of the quark zero
modes in a random instanton vacuum~\cite{INSTREV}, by which they were
originally inspired. 

%%%%%%%%%%%%%%%%%%%%%%%%%%%%%%%%%%%%%%%%%%%%%%
%%%%%%%%%%%%%%I AM HERE%%%%%%%%%%%%%%%%%%%%%%%%
%%%%%%%%%%%%%%%%%%%%%%%%%%%%%%%%%%%%%%%%%%%%%%

\subsection{Microscopic universality in vacuum}
 
The fact that the spontaneous breakdown of chiral symmetry 
along with its explicit breaking  provides powerful constraints
on on-shell amplitudes is widely known and forms the basis
of chiral perturbation theory~\cite{GASSERLEUT1}. 
What is less known is that the same mechanism provides a severe
constraints in the microscopic regime. To show this,
consider the pion  propagator in a periodic Euclidean box,
\be
G_{\pi}(x)=\frac{1}{V}\sum_{K_n} \frac{e^{-iK_n x}}{K_n^2+m_{\pi}^2}
=\frac{1}{Vm_{\pi^2}}+
\frac{1}{V}\sum'_{K_n} \frac{e^{-iK_n x}}{K_n^2+m_{\pi}^2}
\label{pion}
\ee
where the primed sum is over non-zero modes $2\pi n_i/L$, $(i=1,2,3,4)$.
In the macroscopic regime $mV\gg1$,  Gell-Mann-Oakes-Renner relation
$m_{\pi}^2=m \Sigma/F^2_{\pi}$ (where  $\Sigma=|\corr{\bar{q}q}|$) 
allows the usual chiral power
counting,
leading to standard chiral perturbation theory.

In the microscopic regime $mV\ll1$, the primed sum is negligible,
since 
the zero mode contribution $(mV)^{-1}$ dominates (\ref{pion}). 
The chiral counting breaks down.  However, it is still possible  
to resum the zero modes. The general result, obtained by Gasser and 
Leutwyler~\cite{GASSERLEUT2} could be rewritten in terms of partition function:
\be
Z(m)=\int dU e^{V\Sigma(mU+h.c.)}
\ee
Note that partition function does not involve any derivatives.
The partition function depends solely on:\\
(i) - The nature of the coset (pattern of spontaneous breakdown of chiral
symmetry), hidden in the integration measure.\\
(ii) - Explicit breaking of chiral symmetry ( representation $(N_f,N_f)$.)\\

Next important observation, due to Leutwyler and Smilga~\cite{LEUTSM}, 
 is based on the following transformation based on the $\theta$ angle
\be 
Z_n(m) =\int_0^{2\pi} \frac{d \theta}{2\pi} e^{-in\theta}Z(me^{i\theta
/N_f})= \int dU (\det U)^n e^{V(mU+h.c.)}
\label{LS}
\ee
L.h.s., by definition, 
 describes the QCD partition function
 for fixed topological (winding) number $n$.
R.h.s. is rephrased in hadronic variables.
Comparing the same powers of $m$ on both sides yields an infinite 
hierarchy of sum rules, relating the eigenvalues coming from quark
determinant in $Z_n(m)$ to hadronic parameters on the r.h.s. of (\ref{LS}). 
For example, comparison of quadratic  terms 
leads to 
\be
\frac{1}{V^2}\Big\la\sum'_k
\frac{1}{\lambda^2_k}\Big\rangle_n=\frac{\Sigma}{4(n+N_f)}
\ee
It was conjectured by Shuryak and Verbaarschot~\cite{SHURVER}, that there must exist 
a master formula, based on  some microscopic spectral density $\nu_s(x)$, 
where $x= V\lambda$, 
which generates, via taking suitable moments, the LS sum rules. 

At this step the impatient reader may ask, 
how do the random matrices come into 
this scenario?
The answer is simple - let us construct the simplest (minimal) model
which satisfies both constraints (i) and (ii). For pedagogical reason
 let us 
constrain temporarily to one flavor. Such a model is a four-fermi model
in 0 dimensions:
\be
L=q^{\dagger} im e^{i\gamma_5 \theta} q + \frac{1}{V}
q^{\dagger}_Rq_L
q_R^{\dagger}q_L
\label{NJL}
\ee
where the dimension of Grassmann fields is
$n_L$, ($ n_R$) for ``left'' (``right'') components, respectively, 
analog of $\gamma_5$ is defined as $(\bf{1}_{n_R},-\bf{1}_{n_L})$
and the ``volume V'' of the system is nothing but $n_L+n_R$. 
It is not surprising at all, that such a model is consistent with 
restrictions of chiral symmetry breaking, since, 
it is just the zero mode  sector of 
usual Nambu-Jona-Lasinio model!

What is perhaps more surprising, is that this model can be rewritten 
in terms of the bosonic variables $A^{ab}=q_l^a q_R^{b \,\dagger}$ in the
following
form~\cite{BOOK} 
\be
Z(\theta,m)=\int dA e^{-V {\rm tr} AA^{\dagger}} \det
\left( \begin{array}{cc} ime^{i\theta} &
A \\ A^{\dagger} & ime^{-i\theta} \end{array} \right)
\label{RMM}
\ee
But this is, by definition, 
 the partition function for the \underline{chiral} ({\it cf.} block
structure),
\underline{random} Gaussian ({\it cf.} quadratic potential
 in the exponent) {\underline{matrix} ({\it cf.}
matrix-valued field A) model.

This identification of the class of random matrix models 
allows  to use the whole arsenal of RMM to get the
microscopic spectral density. Using the variant of orthogonal
polynomial
method, Verbaarschot and Zahed~\cite{VERZAHED}
 found an explicit form of microscopic 
spectral density for massless QCD
\be
\nu_s(x)=\frac{1}{2}\Sigma^2x(J^2_{N_f}(\Sigma x) - J^2_{N_f+1}(\Sigma x)
J^2_{N_f-1}(\Sigma x))
\label{JI}
\ee

It is a straightforward exercise to show, that the moments of 
microscopic spectral density generate, via doable integrals 
over Bessel functions, the r.h.s. of the Leutwyler-Smilga sum rules.

The last unresolved issue is the universality, i.e. the independence
of 
the microscopic spectral density on the form of the matrix potential
 (one could easily visualize higher powers of $AA^{\dagger}$ in the
exponent). 
At first sight, such universality looks like a miracle, and only very 
recently was proven to hold for arbitrary series-like potentials 
by Damgaard and collaborators~\cite{PAUL1} using rather
formal methods of RMM.  Actually, the physical reason of universality 
is simple - all higher order than quartic terms in NJL-like lagrangian
are suppressed in the thermodynamical   limit (i.e. 
behave like 
$1/\sqrt{V}$).
Let us finally mention, that sum rules and microscopic spectral
densities
were successfully checked in several models of the QCD vacuum, 
e.g. instanton liquid model~\cite{VERACTA}.  They were also checked with high
accuracy
on the lattice~\cite{TILO1}, albeit in the last case the form of microscopic
 universality is of different type (for $SU(3)_c$ Kogut-Susskind
fermions),
 due to
the 
known ambiguities caused by Nielsen-Ninomiya theorem (fermion doubling).

\subsection{Microscopic universality  in matter}

We may ask, to what extent the microscopic universality
survives the effects of matter, e.g. the temperature or presence of
the 
baryonic
potential.
Generic arguments from previous section based on power counting  
 suggest that most of the observations survive the changes in 
external, deterministic parameters, provided that chiral symmetry
is still broken. To make this argument more transparent, let us bring
the standard universality argument by Pisarski and
Wilczek~\cite{PISARSKIW}
 based
on mapping of the QCD onto the linear sigma model. 
For exemplary case of two-flavor QCD in the vicinity of phase transition, 
the $\sigma$ model takes  the form
\be
L = Tr (m\Phi+ h.c.) + g_0(T) {\rm Tr} \Phi^{\dagger} \Phi +\cdots
\label{PW}
\ee
with $g_0(T)$ proportional to the reduced temperature.
The only effect for the crucial for LS-sum rules  first term
in  this Lagrangian is to ``dress'' thermally 
 constant modes: $\Phi \rightarrow \Phi(T)$. 
We therefore suspect that all LS sum rules and microscopic
spectral densities  will stay unchanged, 
provided we replace the scale $\Sigma$ of RMM by $\Sigma(T)$.
The explicit form of temperature dependence of the scale is not
universal,
 i.e. 
depends on the potential of RMM. Here, equivalently, depends
on the explicit form of relevant operators (dotted terms).
This generic scenario agrees with detailed calculations~\cite{TILO2}.
Similar arguments, modulo subtleties related to the formation of the
nucleon Fermi surface, could be presented in the case of baryonic
potential.

Does it mean that the microscopic spectral formula is always sacred?
Not necessarily. It was pointed in~\cite{JNZ} that when 
one also rescales the masses, i.e. in the limit $y\equiv mV \sim
\lambda V=x$, microscopic spectral density will become 
function of two variables, i.e. $\nu_S(x,y)$, provided 
unquenching of the fermionic determinant. 
The universality of these ``double microscopic'' spectral densities
was  proven by Damgaard et al.~\cite{PAUL2}
Despite the double limit  procedure is not physical,
the new sum rules generated by double spectral densities 
could serve as a theoretical concept allowing
better assessment of spontaneous breakdown of chiral symmetry on the
lattice.

\subsection{Structural changes of Dirac spectrum - temperature}

In this subsection we will try to get some insight how 
external parameters may influence the spectrum of random matrix
model. 
The influence on microscopic features was discussed above. 
Here we would like to model the non-universal ({\it cf.}  $\Phi(T)$), 
albeit 
perhaps qualitatively correct thermal behavior of quark condensate.
Let us start again from Banks-Casher relation. 
For zero temperature it is easy to check, that indeed the average
spectral density of RMM is non zero for $\lambda=0$. 
We expect it intuitively anyhow, since we know that RMM is just the 
longest wavelength limit of NJL model, but let us recover this result 
using RMM point of view. 

In practice, it is more convenient to work with Green's functions
then spectral densities. Let us therefore define the Green's function 
for Gaussian random matrix model (for $\theta=0$):
\be
G(z)=\frac{1}{2N} \left< \frac{1}{z-
\left( \begin{array}{cc}
        m & iA \\ iA^{\dagger} & m
        \end{array} \right)
} \right> \equiv  \frac{1}{2N} \left< \frac{1}{z-{\bf Q}} \right>
\label{greenrec}
\ee
For our purposes (zero virtuality) 
 we need to calculate this resolvent at $z=0$ only, or 
equivalently, for any $z$ with $m=0$ and finally put $z=im$. 
Here matrices A are $n_R \times n_L$.  
The two diagonal 
blocks have implicit size $n_{R/L} \times n_{R/L}$.
For fixed $n_{R/L}$, we use the chiral basis
\be
G=g {\bf 1} + g_5 \gamma_5 
\ee
with  ${\bf 1} \equiv {\rm diag} ({\bf 1}_{n_R},
{\bf 1}_{n_L}) $  therefore, after performing the averaging 
$\la..\rangle$  with Gaussian measure,  
we get chirality even and chirality  odd resolvents
\be
{\rm Tr }G&=&g+xg_5 = \frac{1}{2}(z-\sqrt{z^2-4+4x^2/z^2})|_{z\equiv im}
\nonumber \\
{\rm Tr }G\gamma_5&=& xg+g_5 =-\frac{x}{z}|_{z\equiv im}
\label{both}
\ee
Since spectral densities are just related to imaginary 
part   of the Green's
functions\footnote{Note $1/(y+i0)=PV 1/y -i\pi \delta(y)$.}, 
we immediately read out relevant spectral densities
\be
\nu_+(\lambda, x)&=&|x|\delta(\lambda) +\frac{1}{2\pi |\lambda|}
\sqrt{(\lambda^2-\lambda^2_-)(\lambda^2-\lambda^2_+)} \nonumber \\
\nu_-(\lambda, x)&=&x\delta(\lambda)
\label{spectralboth}
\ee
with $\lambda^2_{\pm}= 2\pm2\sqrt{1-x^2}$ and asymmetry  $x=(n_R -n_L)/2N$ . 
Note that for $x=0$ one recovers Wigner's semicircle
law~\cite{WIGNER}.
 In particular, 
the chirality even spectral function $\nu_+$ 
does not vanish at $\lambda=0$, leading therefore,
via analogy to Banks-Casher relation, to ``quark condensate''.

The chirality odd part $\nu_-$ is a direct measurement of the difference
in the spectral distribution between left-handed and right-handed ``quark
zero
modes''. If resembles naively Atiyah-Singer theorem, despite here  the origin 
of the asymmetry  is purely kinematical (``rectangularity'' of the ensemble),
 and not related to any topology. We will come back to this point when
discussing the $U(1)$ anomaly, so in the mean-time let us put $x=0$. 

Now we see how to introduce external parameters into the random matrix 
model - all we have
to do is to bosonize NJL model in matter and look at the softest
modes.
Therefore the modifications of the medium correspond to 
to the modification of ${\bf Q}$ 
 by 
\eq
\arr{0}{i{\bf \Omega}-\mu}{i{\bf \Omega} +\mu}{0}+\arr{m}{iA}{iA^{\dagger}}{m}
\label{repla}
\eqx
where $\mu$ is the chemical potential for ``quarks'' and 
${\bf \Omega}=\omega_n {\bf 1}_n \otimes {\bf 1}_N
$. Here $\omega_n$  are all fermionic Matsubara 
frequencies. 
For simplicity, we restrict ourselves to the lowest pair of Matsubara
 frequencies $\pm\pi 
T$ and the chiral limit. This corresponds to the
model 
proposed in~\refnote{\cite{JV}}.
To investigate chiral symmetry breaking in this
model we should calculate the Green's function and, through
Banks-Casher relation, obtain the condensate. 
We have therefore to ``add'', modulo trivial overall factor $i$, 
 schematic deterministic (D) and random (R) ``Hamiltonians''.
\be
\arr{0}{\pi T}{\pi T}{0} + \arr{0}{A}{A^{\dagger}}{0}
\label{diracT}
\ee
At this moment we could use some of the Blue's magic, here addition
theorem.
 The Blue's function for the model (\ref{diracT})
satisfies (see Appendix)
\eq
B(z)-z=B_D(z)
\eqx
where $B_D$ is the Blue's function of the deterministic
piece. By definition we have
\eq
z-G(z)=B_D(G(z))
\eqx
Now if we evaluate the Green's function of the {\em deterministic}
piece on both sides of this equality we will get so-called
Pastur-Wegner
 equation~\cite{PASTUR}
\eq
G_D(z-G(z))=G_D(B_D(G(z))) \equiv G(z)
\eqx 
The  explicit form of deterministic Green's function is 
$G_D(z)=\frac{1}{2}(\frac{1}{z-\pi T} +\frac{1}{z+\pi T })$, 
since the deterministic eigenvalues are $\pm
 \pi T$ (lowest Matsubara frequencies), so the final result reads
\eq
\label{cardano}
G^3-2zG^2+(z^2-\pi^2 T^2+1)G-z=0
\eqx
Figure~1  shows the imaginary part of the relevant 
solution of (\ref{cardano}) as a function of eigenvalues and 
temperature.\footnote{The other two solutions 
are either real or lead to negative spectral density.}
 The structural change at the forking point
corresponds to vanishing $\nu(\lambda=0)$, therefore restoration 
of chiral symmetry.
\begin{figure}
\centerline{\epsfysize=80mm \epsfbox{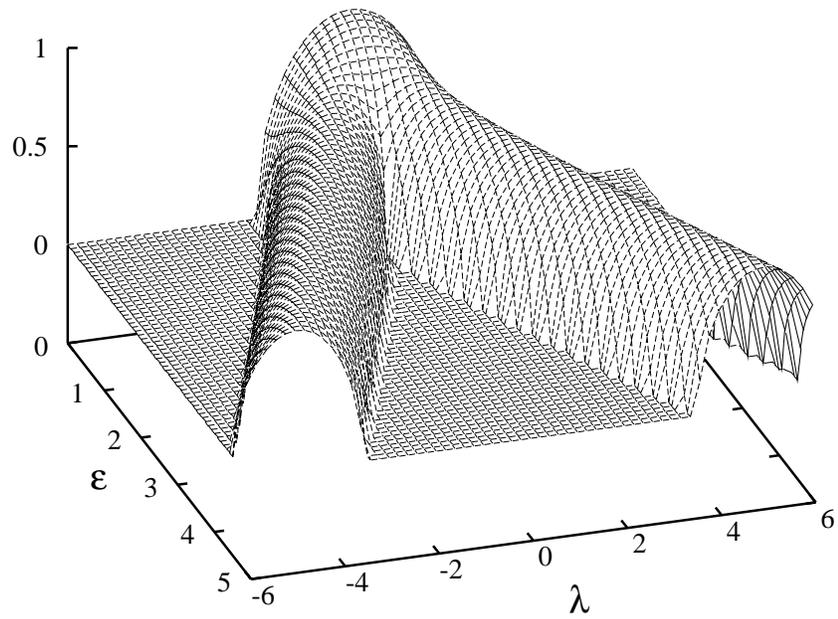}}
\caption{Spectral density $\nu(\lambda)$ as a
 function of deterministic parameter
$\epsilon=\pi T$. 
}
\label{f.spodnie}
\end{figure}

\subsection{Critical temperature}

Critical temperature, corresponding to the point of restoration
of the ``chiral symmetry'' of this model, could be inferred from 
differentiation of the Blue's function corresponding to
(\ref{cardano}).
Intuitively it is obvious  - the endpoints are branching points, 
so the derivative of the Green's functions blows up, hence the
functional 
inverse of Green's function (Blue's function) takes extremum~\cite{ZEE}. 
The condition $B'(G)=0$, together with (\ref{cardano}) leads to the 
real endpoints of the spectra located at
\be
\pm I_1&=&\frac{1}{\sqrt{8}\pi T}
\frac{ (4\pi^2T^2-1+\sqrt{8\pi^2T^2+1})^{\frac{3}{2}}}{\sqrt{8\pi^2T^2
+1}-1 }
\nonumber \\
\pm I_2&=&\frac{1}{\sqrt{8}\pi T}
\frac{(4\pi^2T^2-1-\sqrt{8\pi^2T^2+1})^{\frac{3}{2}}}{\sqrt{8\pi^2T^2
+1}+1}
\ee
The condition $I_2=0$, corresponding to the situation
when two-arc support of the spectrum $[-I_1, -I_2] \cup [I_2,I_1]$
evolves into the single interval on the cut $[-I_1,+I_1]$
corresponds to ``phase transition'' and defines the critical temperature
$T_*=1/\pi$, in units where the width of the random Gaussian
distribution is set to 1.

\subsection{Critical exponents}

The set of all critical exponents could be easily 
inferred from solving algebraic equation (\ref{cardano}) 
and calculating the free energy
of the system through  integrating Blue's functions (see Appendix) 
giving free energy
\eq
\label{free}
F=G^2+\log \f{z-G}{G}
\eqx
Explicitly, they are: ($\alpha, \beta, \gamma, \delta, \nu, \eta)=$
$(0, 1/2, 1, 3, 1/2, 0$), respectively.
 There are mean-field type independently on the number of 
Matsubara frequencies used, in agreement with general 
arguments~\refnote{\cite{BOOK}}.

\subsection{Chemical potential - ``phony vacua''}

Let us now move to the case of finite chemical potential, switching
the temperature off for the clarity of the presentation.
 The matrix
model is now given by
\eq
\label{diracmu}
\arr{0}{-\mu}{\mu}{0}+\arr{0}{A^\dagger}{A}{0}
\eqx
and was originally solved by Stephanov~\cite{STEPHANOV} 
using standard RMM tools. 
The deterministic part is now non-hermitian 
(eigenvalues are complex numbers)and we
have two distinct regions (at least in the quenched case). The
Green's function in the {\it holomorphic region} outside the blob of
eigenvalues is given by (\ref{cardano}) with the substitution $\pi^2
T^2 \ra -\mu^2$:
\eq
\label{cardanomu}
G^3-2zG^2+(z^2+\mu^2+1)G-z=0
\eqx
Instead of following the original argument~\cite{STEPHANOV}, 
we suggest again the use 
of Blue's. 
In order to find the boundary of the nonholomorphic
region we may exploit the method of conformal mapping
 via the Blue's functions (see schematic figure~2), and transform
the cuts of the $T\neq 0$ case into the boundary by the transformation
\eq
z \ra w= z-2G(z)
\eqx
where $G(z)$ is the appropriate branch of the cubic equation
(\ref{cardano}) with {\it formal} identification $\pi^2T^2=+\mu^2$. 

\begin{figure}
\label{f.map}
\centerline{\epsfysize=60mm \epsfbox{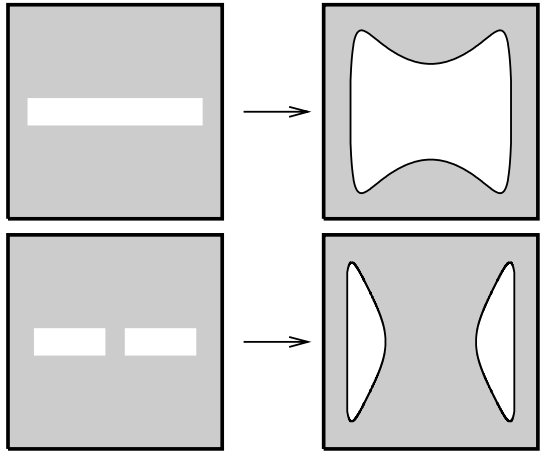}}
\caption{Schematic mapping illustrating conformal 
transformation of the cuts supporting the eigenvalue 
distribution visible at figure 1 onto the boundaries of islands
for complex eigenvalue distribution.
The shaded regions represent holomorphic domain, where the mapping is
valid.
}
\end{figure}

The result of this mapping (continuous line) is presented on Fig. 3.
The dots are the numerically generated eigenvalues of (\ref{diracmu}).
The mapping reproduces immediately results of original random matrix model 
for $\mu$~\refnote{\cite{STEPHANOV}}. The islands reproduce the domains
of mixed-quark condensate (``phony vacua''), 
obtained as an artifact of neglecting in the process of averaging the
phase 
of the complex fermion determinant. 
This condensate is not the usual chiral quark condensate, 
but quark-conjugate quark condensate which forms due to the fact
that quenching of the phase stops penalizing attraction leading to such 
configurations. Such a condensate violates spontaneously baryon
number.\footnote{
Vafa-Witten restrictions do not hold for complex measure.} 
This model suggests the reasons for 
failure 
of the quenched lattice calculations with chemical potential, due to
the
appearance of this mixed-condensate or, equivalently, infinitely growing
 fluctuations 
when approaching the  boundaries
of the islands~\refnote{\cite{USMUX}}, due to the appearance of
so-called
``baryonic pion''. 

\begin{figure}
\label{f.chem}
\centerline{\epsfxsize=10cm \epsfbox{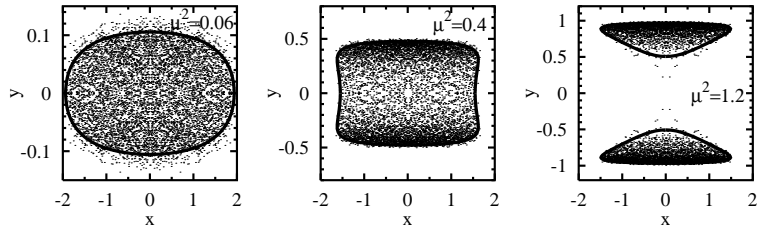}}
\caption{Comparison of the numerically obtained distribution of
eigenvalues
for few sample chemical potentials with analytical results.
The solid lines represent the boundaries obtained by conformal mapping.
}
\end{figure}

\subsection{Chemical potential - ordinary vacua}

We may also consider the random model without suppressing the phase of
the determinant.
The investigation of the behavior of unquenched partition function
(e.g. location and densities of  Lee-Yang zeroes~\cite{HUANG}) 
could be immediately inferred
from integration of Blue's functions. The location of
singularities separating the ``phases'' of the system
may be found  from equating the real parts of the free
energies  calculated for different solutions (labeled $(i,j)$) of the
cubic Pastur-Wegner  equation (\ref{cardanomu}).
Explicitly, 
the location of singularities of partition function for model 
(\ref{diracmu}) is given by 
\be 
{\rm Re }F_i={\rm Re }(G_i^2+\log \f{z-G_i}{G_i})
 =
{\rm Re }(G_j^2+\log \f{z-G_j}{G_j})={\rm Re }F_j 
\ee
 This procedure describes very
nicely   precise (up to 500 digit accuracy) 
numerical calculations of \cite{HALASZ}.
The resulting phase structure suggests first order
phase transition, as discussed in~\cite{BOOK}.

\subsection{Phase diagram}

The combination of temperature and chemical potential can be easily
investigated as discussed in subsection 2.4.
The $1/N$ approximation is reminiscent of the mean-field 
treatment  discussed originally in~\cite{BOOK}.
The phase diagram (for chiral case) for condensate is plotted on
Fig.~4. 
For chemical potential $\mu > \mu_{crit}=1/6\Sigma \ln (2+\sqrt{3})$
the transition is first order (due to lack of vector couplings
in 0d NJL model) and disappears at $\mu=0.5\Sigma$.
At zero chemical potential, the transition is second order
(mean-field)
with critical temperature $T_{crit}=1/4\Sigma$. The different
numerical 
coefficient comparing to subsection 2.4 stems from the fact, that here
we have taken into account all Matsubara modes. 
\begin{figure}
\label{f.ph}
\centerline{\epsfxsize=10cm \epsfbox{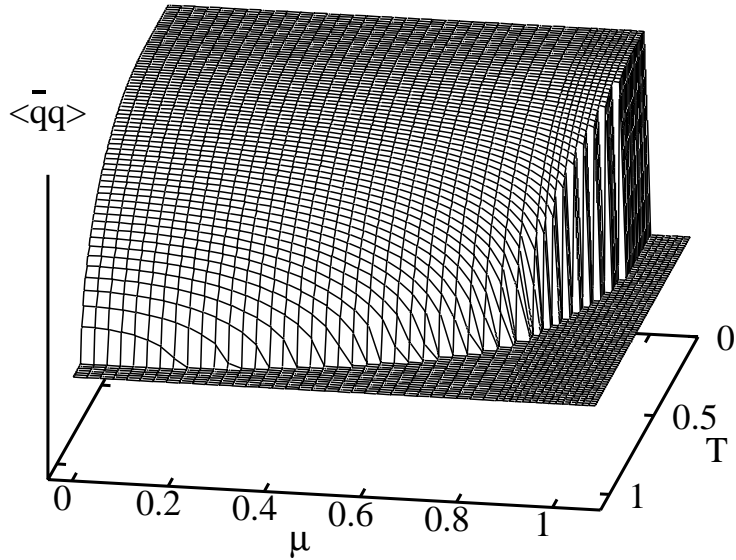}}
\caption{Condensate vs temperature and chemical potential 
for chiral random matrix model with infinitely many Matsubara
frequencies 
included.
}
\end{figure}
It is interesting to compare Fig.~4, obtained using only the constant
modes, to the full NJL result of Weise and Lutz \cite{WEISE}
and also Zhuang et al.~\cite{ZHUANG}

\subsection{U(1) problem}

Let us now look at more subtle applications of random matrix models,
where usual mean field arguments do not hold and the
dynamical effects of fermionic determinant have to be taken into
account. We start from the  $U(1)$ problem, i.e. the fact that 
pseudoscalar boson is much heavier than other pseudoscalars, 
violating bounds based on chiral symmetry.   
This problem is related to axial anomaly.
In subsection 3 we have encounter the ``kinematical version'' of Atiyah-Singer
theorem. Since in QCD Atiyah-Singer theorem is nothing but 
the global (integrated) axial anomaly, we may hope that the 
effects of fluctuations in $(n_R-n_L)$, leading to 
chirality-odd 
spectral density, may mimic some features of anomalous sector of
 QCD. 

 Naively, it looks that the effects of asymmetry $x$ are meaningless
in the large $N$ limit. Variable $x=(n_R -n_L)/2N$ seems to be irrelevant 
in the large $N$ limit. This reasoning is false, due to the fact, 
that the subleading  at the first glance  terms coming from asymmetry
are enhanced by  factor $N$ coming from fermionic loops.
The resolution of this problem requires therefore an infinite resummation
of contributions originating from fermion determinant.

 Technically, we could do it in one step using the integration
of the Blue's (see Appendix).  For a fixed asymmetry 
$x\equiv \chi/2N$ we have
\be
\partial_z \log Z(\chi)=-2NN_fi {\rm Tr} G
\ee
where ${\rm Tr} G$ is given by first relation in (\ref{both}).
When integrating Blue's or Green's functions
 we need only 
to study the contribution of order $x^2$ from Green's function, 
therefore $G(z)\approx -x^2/(z^2\sqrt{z^2-4})$.
Straightforward integration leads to 
\be
[\log Z(\chi)]_2=\frac{1}{2N}N_f \frac{1}{m(\sqrt{m^4+4}+m)}\chi^2
\label{part2}
\ee
where index reflects the approximation  
and the constant was fixed from imposing the right asymptotic behavior. 
Full partition function is 
\be
Z=\int d\chi e^{\frac{-\chi^2}{2\chi^* N}}Z(\chi)
\ee
where integration over $\chi$ takes into account the 
fluctuations in number of right-handed and left handed zero modes.
The Gaussian measure is consistent with axial anomaly~\cite{ALKOF}. 

 Inserting (\ref{part2}) into above equation leads to 
replacement
\be
1/\chi^* \rightarrow 1/\tilde{\chi}^*=1/\chi^* +\!
\sum_{i=1}^{N_f}\frac{1}{m_f(\sqrt{m_f^4\!+\!4}\!+\!m_f}) \equiv 1/\chi^* 
\!+\!\sum_{i=1}^{N_f} \gamma_i 
\label{screening}
\ee
where we have used the shorter notation
and we have reinstated more general (unequal flavors) flavor dependence. 

Note that for $N_f=0$ (quenched) the topological susceptibility is just 
the given by $\chi^*$, as it should, reminding the Witten-Veneziano
relation known from QCD without light flavors. 
The unquenched result $\tilde{\chi}^*$ shows the screening caused 
by fermion determinant (quark loops) and vanishes in the chiral limit, 
in agreement with the theorems of the QCD.
The formal trick with the integration of the Blue's functions allowed us
to resum infinitely many  infrared terms, and obtained a 
strictly non-perturbative result.
 ``Quark loops'' 
(terms of order $x^2$) from fermion determinant were absolutely
crucial for this resummation, since quenched result does not lead
to the screening. The presented result is the 
analog of the screening phenomenon in a statistical ensemble made of 
negative and positive topological charges in full (unquenched) QCD.
Similar mechanism works in the instanton vacuum~\cite{WEISS}.

\subsection{Pseudoscalar susceptibility}

Let us finally demonstrate that the manifestations of the U(1) anomaly are
identical 
in QCD and our random matrix model, leading to the same generic 
screening of pseudoscalar susceptibility. 
We note that the RMM leads to 
``Ward identity'' relating topological susceptibility, pseudoscalar 
susceptibility and condensate
\be
\tilde{\chi}=-\frac{2m}{N_f^2}i {\rm Tr G}-\frac{m^2}{N_f^2}\chi_{ps},
\ee
in analogy to anomalous Ward identity in Minkowski QCD
\be
i\chi_{top}=-\frac{im}{N^2_f}\corr{\bar{\psi} \psi} 
+\frac{m^2}{N^2_f}\int d^4x \corr{T^* \bar{\psi}i\gamma_5 \psi(x)
\bar{\psi}i\gamma_5 \psi(0)}
\label{wardqcd}
\ee
where the pseudoscalar susceptibility is defined by the integral 
in (\ref{wardqcd}). The resolution of the U(1) problem in QCD stems from
the fact
that for small chiral mass $m$, the absence of the U(1) Goldstone boson
requires that $\chi_{top}=-m\corr{\bar{\psi}\psi}/N_f^2$ 
{\it to order} ${\cal O}(m^2)$, or in other words, pseudoscalar
susceptibility
has to be finite in the chiral limit.  
In random matrix model, the above {\it unquenched} 
calculations and Ward identity 
yield the pseudoscalar susceptibility to be equal~\cite{USU1} 
\be
\chi_{ps} &\equiv & \frac{1}{N} [-\corr{{\rm tr} {\bf Q}^{-1}\gamma_5 
{\bf Q}^{-1}\gamma_5}
+\frac{1}{N} \corr{{\rm tr} {\bf Q}^{-1}\gamma_5{\rm tr} 
	{\bf Q}^{-1}\gamma_5}_{connected}
\nonumber \\
&\sim &(\sum_i^{N_f}\frac{1}{m_i})^2\cdot \frac{1}{1/\chi^*
+\sum_i^{N_f}\gamma_i} -4\sum_i^{N_f}\gamma_i
\label{nontrivial}
\ee
therefore {\it finite} in the chiral limit.
The first equality is the definition of pseudoscalar susceptibility
in chiral random model. 
The second equality is the result of our 
calculations. Note the non-trivial cancellation 
of the infrared singularities $1/m^2$ between the two terms 
in (\ref{nontrivial}). This remarkable
cancellation
is the result of competition between the trace and trace-trace terms in 
(\ref{nontrivial}).
 {\it Naive} expectation based upon large N counting suggests {\it wrongly}
that the trace-trace term is subleading. 
Neglecting the trace-trace term leads to diverging pseudoscalar 
susceptibility in the chiral limit, therefore to   
the ``Goldstone pole'' in U(1) channel, and consequently 
to the failure of the resolution of  the anomaly problem. 
 Like in real QCD, the unquenched treatment of 
large oscillations (${\cal O} (N^{1/2})$ , where N commensurates with
volume)
 in
topological charge are needed for proper physics in the U(1) channel. 
We  found rather remarkable that the simple matrix model  without any
further
dynamical assumptions is able to illustrate the mechanism of screening
consistent with our knowledge of the QCD. 
We do hope also that some of these results are relevant for lattice
simulation
for U(1) observables, taking into account that 
the resummation of infrared
terms might be troublesome on the finite lattice.

\subsection{Finite $\theta$}
QCD at finite vacuum angle $\theta$ is not well understood. Despite
experimental evidence suggest that $\theta <10^{-10}$, the issue of 
potential CP breakdown of QCD poses a serious theoretical challenge. 
Since due to the U(1) anomaly, the $\theta$ term may be traded from gauge
fields to the quark mass matrix, some problems of $\theta$ term 
may be investigated using the chiral random matrix models, 
provided that the mass matrix ${\rm diag}\,(m_1, m_2,...m_{N_f})$ 
is replaced by  $e^{i\theta/N_f}{\rm diag}\,(m_1, m_2,...m_{N_f})$.

Saddle point analysis~\cite{USTHETA} leads to the following unsubtracted free
energy (for small masses)
\be
F&=&\sum_{j=1}^{N_f}\left(p_j^2-\log{p_j^2} 
	-2\frac{m_j}{p_j} \cos {\phi_j}\right) 
\label{theta1}
\\
&+&\frac{\chi_* V}{2N}
	\left(\theta-\sum_{j=1}^{N_f} \phi_j-\sum_{j=1}^{N_f}
\frac{m_j}{p_j} \sin \phi_j\right)^2+O(m_j^2) \nonumber
\ee
where saddle point solutions of RMM are parameterized as
$p_je^{i(\theta/N_f
-\phi_j)}$. Remarkably, this RMM result reproduces Witten 
equations~\cite{WITTEN},
derived using different arguments. For $N_f=3$  the solution at
$\theta=\pi$ is doubly degenerated, implying spontaneous breakdown 
of CP, provided the Dashen bound holds. For physical masses, however,
spontaneous breakdown of CP seems to be excluded.

 Encouraged by these analytical results, we may probe numerically
the {\it
unquenched} $\theta$ problem in the framework of random matrix models.
 At the  left side of Fig.~\ref{f.theta}
we plot the numerical results~\cite{USTHETA} for free energy as a function of 
vacuum angle. On the right side, we compare the similar
plot from~\cite{SCHIERHOLZ}, obtained for dynamical $CP^3$ model (plots
for QCD do not exist due to the complexity of the  problem). 
Both plots exhibit a kink at some critical value, suggesting first
order ``deconfining'' phase transition.  If this 
 transition would persist for
QCD, such picture may suggest a dynamical resolution of the $\theta$
problem. Namely, deconfinement may always happen
due to the fact that chromomagnetic monopoles would acquire 
electric charges $\theta/2\pi$, and would screen in consequence long-range
color forces. Therefore the presence of confining phase
 forces $\theta=0$. 

Such scenario requires that the kink visible on both figures
migrates to zero in the thermodynamical limit.

\begin{figure}
\label{f.theta}
\centerline{\epsfxsize=40mm \epsfbox{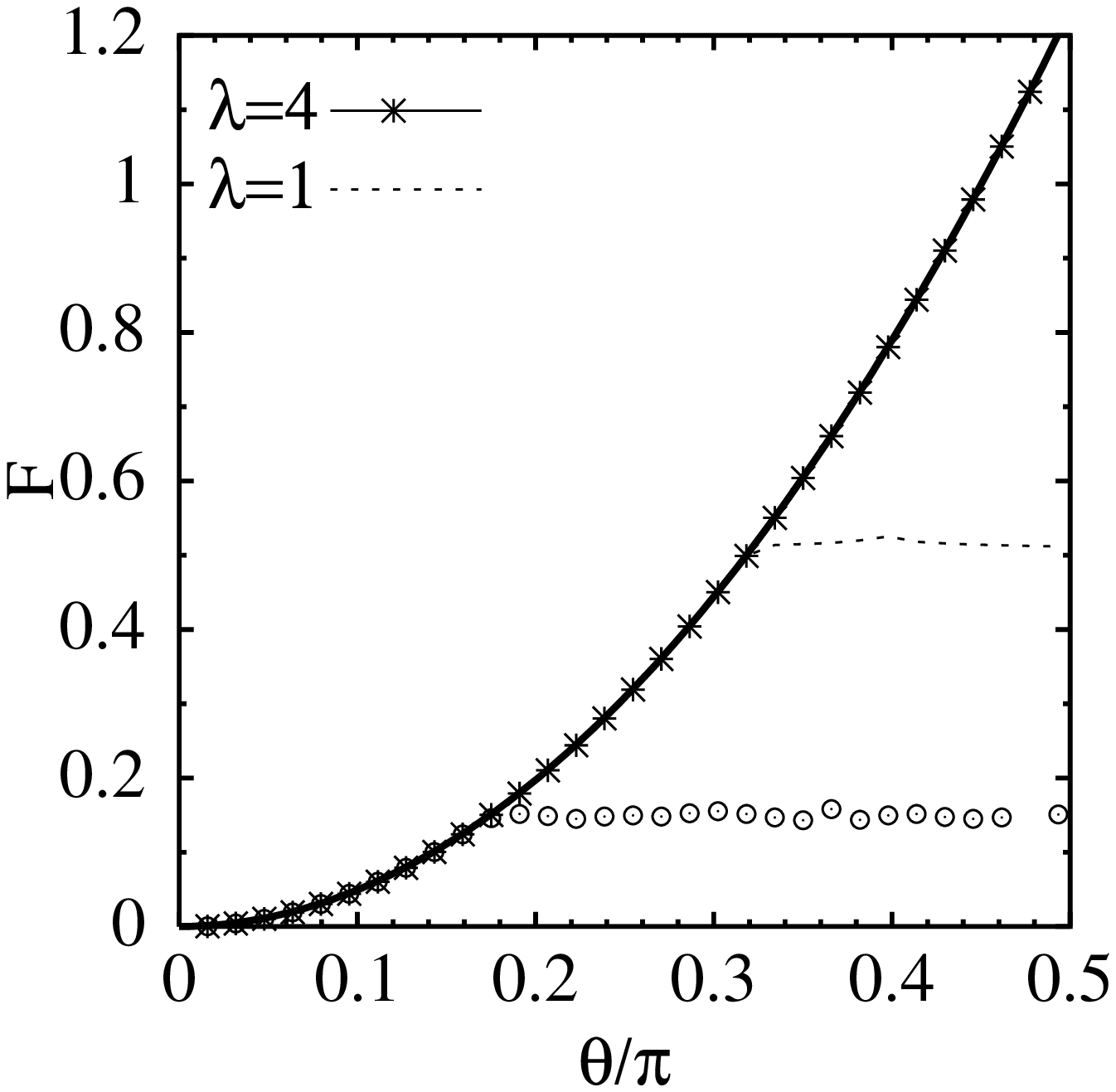} \hspace*{3mm}
 \epsfxsize=42mm \epsfbox{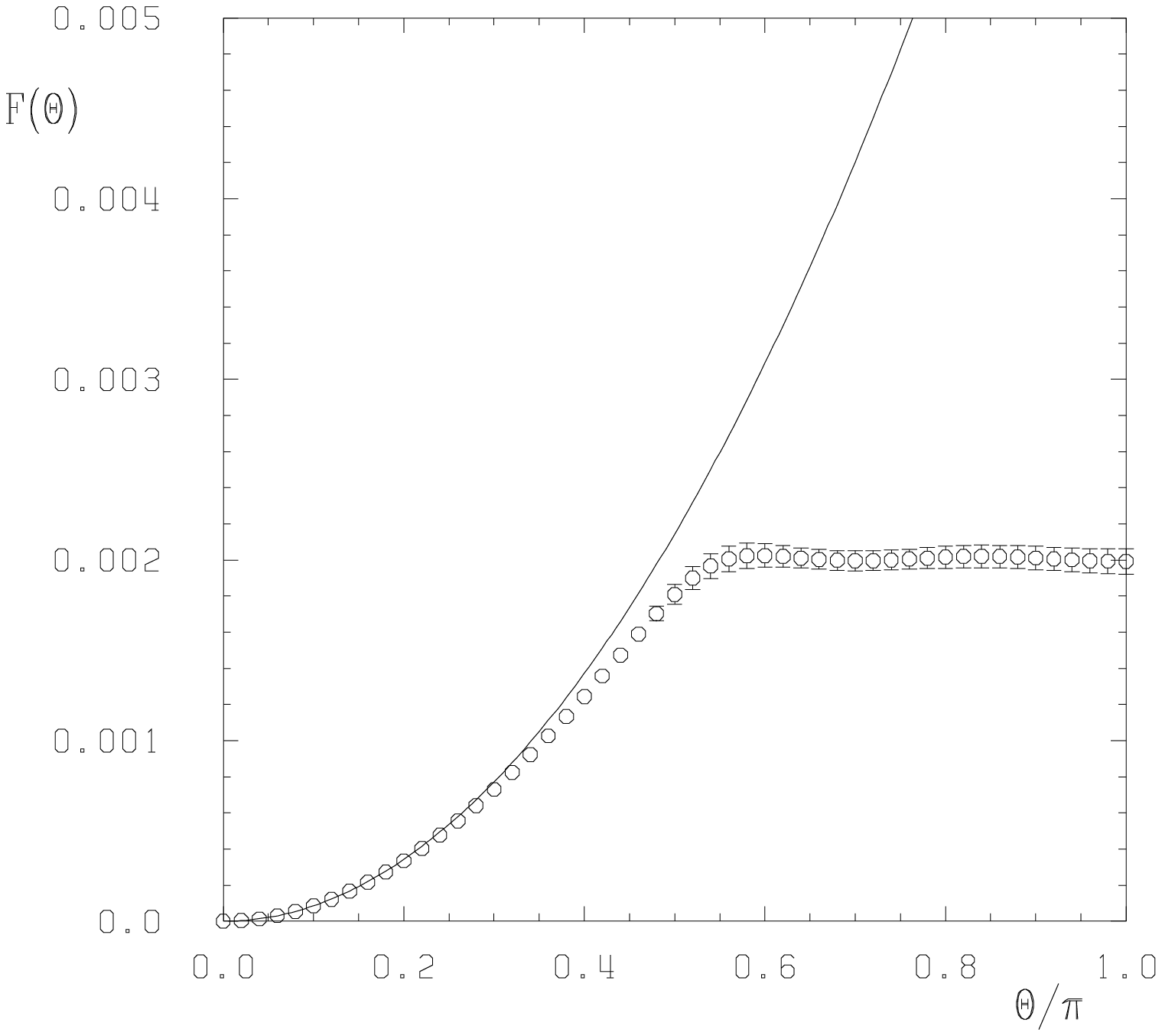}}
\caption{Free energy as a function of $\theta/\pi$ 
for RMM (left) and $CP^3$ model~\protect\cite{SCHIERHOLZ}. Here
$\lambda=N/V$.
}
\end{figure}

RMM allows another scenario consistent with  the CP$^3$ data.
  High precision numerical studies in RMM~\cite{USTHETA}
strongly suggest that the  major factor driving the position of the kink 
seems to be rather the ratio $N/V$, where $N$ refers to number of zero
modes in a fixed volume $V$. Therefore the kink 
observed for dynamical models like $CP^3$ 
{\it may be} the result of subtle
interplay between quark masses, finite size effects and accuracy in
numerical 
assessments.

\section{CONCLUSIONS}

In this talk, we tried to give a quick overview 
 of applications of RMM
to various domains of QCD, with the emphasis on the phenomenon 
of spontaneous breakdown of chiral symmetry.   The main intention
of this presentation was to exhibit the generic behavior based
on fundamental physical premises.     We  therefore have 
skipped several important
technical
details and  we have often referred to original papers.
Meanwhile, we tried to 
 stress  the fact, 
that several ``miraculous'' at the first glance agreements 
of RMM and QCD, either at quantitative level (microscopic regime) or 
at qualitative level  (macroscopic regime), are based
on firm and  well understood principles of spontaneous breakdown of chiral
 symmetry. 

We do hope that this way of looking at RMM in reference
 to QCD may  be helpful at least for a non-expert  in Random Matrix Models.

\section*{Acknowledgments}

MAN thanks the organizers of the Workshop for their hospitality and
for managing to assemble an unusually stimulating set of lectures 
 and discussions.
This work was partially supported 
by the Polish Government Project (KBN) grants 2P03B04412 and 2P03B00814,
by the US DOE grant DE-FG-88ER40388
and by Hungarian grants OTKA T022931 and
FKFP0126/1997.

\section*{APPENDIX - FIVE FACETS  OF BLUE'S}

\newcommand{\sg}{\sigma}
\newcommand{\PP}{{\cal P}}
\newcommand{\dz}{\partial_z}
\newcommand{\dzb}{\partial_{\zb}}
\renewcommand{\gg}{\GG\otimes \GG^T \cdot \Gamma}
\newcommand{\ggb}{\GG\otimes \br{\GG}^T \cdot \Gamma}
\newcommand{\gbgb}{\br{\GG}\otimes \br{\GG}^T \cdot \Gamma}
\newcommand{\ggg}{\det(1-\GG\otimes \br{\GG}^T \cdot \Gamma)}

The fundamental problem in random matrix theories is to find the 
distribution of eigenvalues in the large $N$ (size of the matrix $\MM$)
limit~\cite{REVIEWMAT}. One encounters two generic situations --- in the case of
hermitian matrices the eigenvalues lie on one or more intervals on the real
axis, while for general non-hermitian ones, the eigenvalues occupy
two-dimensional domains in the complex plane. In both cases the
distribution of eigenvalues 
can be reconstructed from the knowledge of the Green's function:
%The eigenvalue distribution is easy to reconstruct from the
%discontinuities
%of the Green's function 
\be
G(z)=\frac{1}{N}\left< {\rm  tr} \frac{1}{z-\MM}\right>
\label{green}
\ee
where averaging is done over the ensemble of $N \times N$ random matrices
generated with probability
\be
P(\MM)=\frac{1}{Z}e^{-N {\rm tr } V(\MM)}
\ee
For hermitian matrices the discontinuities of the Green's function
coincide with the eigenvalue supports.
In the simplest example of the Gaussian potential $V(\MM)= \MM^2$ for 
{\it hermitian} matrix $\MM$, a standard calculation gives
\be
G(z)=\frac{1}{2}(z-\sqrt{z^2-4})
\label{gauss}
\ee
The reconstruction of the spectral function ({\it i.e.} 
eigenvalue distribution)
is based on the well known relation
\be
\frac{1}{x\pm i\epsilon}= {\rm PV} \frac{1}{x} \mp i\pi \delta(x)
\ee
Then, the spectral function turns out to be 
\be
\nu(\lambda)= -\frac{1}{\pi} \lim_{\epsilon \rightarrow 0} {\rm Im} 
G(z)|_{z=\lambda +i\epsilon}
\ee
In the case of the Gaussian Green's function (\ref{gauss}), the
discontinuities in the 
Green's function come from the cut of the square root, leading to the 
Wigner's semicircle law $\nu(\lambda)=\frac{1}{2\pi}\sqrt{4-\lambda^2}$
for the eigenvalue distribution of random hermitian matrices.

The Blue's function is defined as a functional inverse of the Green's 
function
\be
B[G(z)]=z
\label{blueinv}
\ee
For random Gaussian ensemble, the functional inverse of (\ref{gauss})
is simply $B(z)=z+1/z$. 
Blue's functions were  introduced by Zee~\cite{ZEE} as the result 
of diagrammatic reinterpretation of the seminal addition formalism
(R-transformation) by Voiculescu~\cite{VOICULESCU}.

\subsection{Addition}

Let us consider the problem of calculating Green's function 
for the sum of two independent ensembles $\MM_1$ and $\MM_2$, 
i.e.
\be
G(z)&=&
\frac{1}{N} \int [d\MM_1][d\MM_2] P(\MM_1)P(\MM_2)
{\rm tr} \frac{1}{z-\MM_1-\MM_2} 
\label{conv} \\
&\equiv&\frac{1}{N}\left< {\rm tr} \frac{1}{z-\MM_1-\MM_2}\right>
\nonumber
\ee
The concept of addition law for hermitian ensembles 
relies on 
 additive transformation, which linearizes
the convolution of non-commutative matrices (\ref{conv}), alike to the 
logarithm of the Fourier transformation  for the convolution 
of  arbitrary functions. This method is 
 an important shortcut to obtain the equations for the Green's
functions
for a sum of matrices, starting from the knowledge of the Green's functions
of individual ensembles of matrices.

The addition law for Blue's functions reads~\cite{ZEE}
\be
B_{1+2}(z)=B_1(z)+B_2(z)-\frac{1}{z} .
\label{addblue}
\ee
Using the definition of the Blue's function we could rewrite the last
equation
in the ``operational'' form
\be
z=B_1(G)+B_2(G)-G^{-1}
\label{addbluebis}
\ee
with $G_{1+2}\equiv G$. 
The algorithm of addition is now surprisingly simple~\refnote{\cite{ZEE}}:
Knowing $G_1$ and $G_2$, we find (\ref{blueinv}) $B_1$ and $B_2$.
Then we read out from (\ref{addbluebis}) the final equation for the 
resolvent for the sum.
Note that the method treats on equal footing the Gaussian and
non-Gaussian 
ensembles, provided that the ensembles are sufficiently independent
(free).
It is also applicable for chiral random matrices. 
For generalization of this algorithm for the case arbitrary complex
random matrices we refer to original papers~\cite{USDIAG,ZEEFE,SHADES}.

\subsection{Multiplication}
Blue's functions provide also an important shortcut
to obtain the equation for the Green's (Blue's) function 
for a product of matrices, starting from knowledge of Green's (Blue's) 
functions
of individual ensembles of matrices, i.e. to find  
\be
G(z)=\frac{1}{N}\left< {\rm tr} \frac{1}{z-\MM_1 \cdot \MM_2}\right>   
\label{product}
\ee
provided that $\MM_1$ and $\MM_2$ are free and $\corr{{\rm tr}\MM_1},
 \corr{{\rm tr}\MM_2}
\neq 0$. 
Introducing the notation for continued fraction
\be
\frac{w}{\Sigma_{\MM}\left( \frac{w}{\Sigma_{\MM}\left(
 \frac{w}{\Sigma_{\MM}(...)}\right)}\right)}  
\equiv   \frac{w}{\Sigma_{\MM}(\bullet)}
\label{bullet}
\ee
where self-energy $\Sigma(z)$ is defined in a standard way
$1/G(z)=z-\Sigma(z)$
multiplication law for Blue's functions reads\refnote{\cite{JANIKPHD}}
\be
B_{1\star
2}\left(\frac{w}{\Sigma_{1\star2}(\bullet)}\right)
=
\frac{w}{1+w}
B_1\left(\frac{w}{\Sigma_1(\bullet)}\right) \cdot 
B_2\left(\frac{w}{\Sigma_2(\bullet)}\right)
\label{prodblue}
\ee
The usage of Blue's functions
provides the factorization mechanism for averaging 
procedure in (\ref{product}) and is equivalent to S transformation 
of Voiculescu 
 via~\refnote{\cite{JANIKPHD}}
 \be 
S(w)=\frac{1}{\Sigma\left(\frac{w}{\Sigma(\bullet)}\right)}
\label{relation}
\ee
where 
$S(w)=\chi(w)(1+1/w)$ with  $\chi(zG(z)-1)=1/z$.

\subsection{Differentiation}

As demonstrated by Zee, differentiation of the Blue's function 
leads to determination of the endpoints of the eigenvalue distribution 
for hermitian random matrices. 
One could easily visualize this point recalling the form of the Green's
function for Gaussian distribution (\ref{gauss}). When approaching the
endpoints
$z=\pm 2$, the imaginary part of Green's function vanishes like
$\sqrt{(2-\lambda)(2+\lambda)}$. Therefore, the endpoints fulfill the equation
$G'(z)|_{z=\pm 2} =\infty$, which may be used as the 
defining equation for {\it unknown} 
endpoints in case of more complicated ensembles.
Since Blue's function is the functional inverse of the Green's function,
the  location of the endpoints could be
inferred from a simpler relation~\cite{ZEE}
\be
\frac{d B(G)}{dG} =0
\label{blueend}
\ee

\subsection{Integration}

Let us consider the partition function
depending  on a parameter $z$:
\eq
Z_N=\corr{\det(z-\MM)}
\label{zpart}
\eqx
For finite $N$ the partition function is a polynomial in $z$ and has
$N$ zeroes (Yang-Lee zeroes):
\eq
Z_N=(z-z_1)(z-z_2)\dots (z-z_N)
\eqx
Taking the logarithm and approximating the density of zeroes by a
continuous distribution we get
\eq
\log\cor{\det(z-\MM)} = \sum_i \log(z-z_i) = \int \rho(z') \log(z-z')
dz'
\label{INTEGRAL}
\eqx
We see that the density of Yang-Lee zeroes can be reconstructed from
the discontinuities of the {\em unquenched} Green's function:
\eq
\label{der}
\partial_z \log\cor{\det(z-\MM)}=\cor{\tr \f{1}{z-\MM} \cdot \det (z-\MM)}
\eqx
using e.g. Gauss law~\refnote~{\cite{VINK}}. For $z$ close to infinity the
unquenched and quenched (\ref{green}) Green's functions
coincide\footnote{In leading order in $1/N$.}. Moreover the unquenched
resolvent is nonsingular configuration by configuration and hence is
holomorphic.
Therefore it is exactly the functional inverse of the ordinary
hermitian Blue's function. In general we may have several branches of
Green's functions, which are determined by different saddle points
contributing to (\ref{zpart}). The discontinuity (cusp) and hence the
location of Yang-Lee zeroes is determined by the possible contribution
of two of them. From a $1/N$ expansion
\eq
\log Z_N=N F_0+F_1+\f{1}{N}F_2+\ldots
\label{YL1}
\eqx
we see that two saddle points may contribute if
\eq
{\rm Re} F_0^{sp.I}={\rm Re} F_0^{sp.II}
\label{cuspline}
\eqx
and $F_0$ is determined by (\ref{der})
\eq
F_0=\int^z dz' G(z')+const
\label{blueint}
\eqx
or equivalently
\be
E_0 =zG - \int\! dG\ B(G) + const
\label{trick}
\ee
after integrating by parts. Note that we have used the fact 
that $z(G) =B(G)$ is just the Blue's 
function~\refnote{\cite{ZEE}} of $G$, 
The integration using Blue's function is
in most cases much simpler than the corresponding direct integration
of $G(z)$.

\renewcommand{\dz}{\partial_z}
\renewcommand{\dzb}{\partial_{\zb}}
\newcommand{\ddx}{\partial_x}
\newcommand{\ddy}{\partial_y}

\subsection{Mapping}

In case of non-hermitian random matrices, eigenvalues are complex. 
The average density of eigenvalues follows from the electrostatic
analogy to two-dimensional Gauss law 
\be
\nu(z,\bar{z})= \frac{1}{\pi}\dzb G(z,\bar{z})
\ee
Contrary to hermitian case, when eigenvalues condense on cuts, 
here they form two-dimensional nonholomorphic (non-analytic)
``islands''.
Surprisingly, using only the analytical structure of the Blue's
functions,
we can find the boundaries of these ``islands'', 
 while staying wholly within the
holomorphic region~\cite{USMAP}.

Let us consider an example. First,  
we consider the case  where a Gaussian random and hermitian matrix
$H$ is added to an arbitrary hermitian matrix $M$, then the case
where a Gaussian random and anti-hermitian matrix $iH$ is added 
to the same  matrix $M$. 
In first case, eigenvalues are on the cuts (they are real), 
in the second case, form an island. In both cases
the regions free of eigenvalues are holomorphic (here fulfill the 
Laplace equation). This implies that there exists a conformal
transformation relating these two domains of analyticity.  
For the case considered here, Blue's functions for both cases are 
related via
\be
B_{iH+M}(u)=B_{H+M}(u)-2u
\label{sumBLet}
\ee
Substituting $u\rightarrow G_{H+M}(z)$ we can rewrite (\ref{sumBLet}) as
\be
B_{iH+M}[G_{H+M}(z)]=z-2G_{H+M}(z) \,.
\label{map1}
\ee
Let $w$ be a point in the complex plane for which
$G_{iH+M}(w)=G_{H+M}(z)$.
Then, using the definition of the Blue's function, we get 
\be 
w=z-2G_{H+M}(z)\,.
\label{map2}
\ee
Equation (\ref{map2}) provides a conformal transformation mapping
the {\it holomorphic} domain of the ensemble $H+M$ ({\it i.e.}
the complex plane $z$ minus cuts) onto the {\it holomorphic} domains of 
the ensemble
$iH+M$, {\it i.e.} the complex plane $w$ minus the ``islands'',  defining 
in this way the 
support of the eigenvalues.
Such construction could be generalized for other non-gaussian 
ensembles, due to the
fact that addition of Blue's functions forms an abelian group.

%\vglue 2.5cm
%\begin{thebibliography}


\section*{References}
\begin{thebibliography}{99}
\bibitem{ZEE}
A. Zee,	
	{\em Nucl. Phys.} {\bf B474}, 726 (1996).

\bibitem{VOICULESCU}
D.V. Voiculescu, 
	{\em Invent. Math. } {\bf 104}, 201 (1991);
D.V. Voiculescu, K.J. Dykema and A. Nica, 
	``Free Random Variables'',
	Am. Math. Soc., Providence, RI, (1992);
for new results see also A. Nica and R. Speicher,
	{\em  Amer. J. Math.}  {\bf 118}, 799 (1996) and references therein.

\bibitem{BANKSCASHER}
T. Banks and A. Casher, 
	{\em Nucl. Phys.} {\bf B169}, 103 (1980).

\bibitem{USDISORDER} 
R.A. Janik, M.A. Nowak, G. Papp and I. Zahed,
	e-print hep-ph/9803289.

\bibitem{INSTREV}
for recent reviews, see T. Sch\"{a}fer, E.V. Shuryak, 
	{\em Rev. Mod. Phys.} {\bf 70}, (1998) 323;
%	e-print hep-ph/9610451;
D. Diakonov, 
	e-print hep-ph/9602375.

\bibitem{GASSERLEUT1}
S. Weinberg, 
	{\em Physica} {\bf 96A}, 327 (1979);
J. Gasser and H. Leytwyler, 
	{\em Ann. Phys.} {\bf 158}, 142 (1984).

\bibitem{GASSERLEUT2}
J. Gasser and H. Leutwyler, 
	{\em Phys. Lett.} {\bf B184}, 142 (1987);
	{\it ibid.} {\bf B188}, 477 (1987); 
	{\em Nucl. Phys.} {\bf B307}, 763 (1988).

\bibitem{LEUTSM}
H. Leutwyler and A. Smilga, 
	{\em Phys. Rev.} {\bf D46}, 5607 (1992).

\bibitem{SHURVER}
E.V. Shuryak and J.J.M. Verbaarschot, 
	{\em Nucl. Phys.} {\bf A560}, 306 (1993).

\bibitem{BOOK}
M.A. Nowak, M. Rho and I. Zahed, 
	{\em Chiral Nuclear Dynamics}, 
	World Scientific, Singapore, 1996.  

\bibitem{VERZAHED}
J.J.M. Verbaarschot and I. Zahed, 
	{\em Phys. Rev. Lett.} {\bf 70}, 3852 (1993).

\bibitem{PAUL1}
G. Akemann, P.H. Damgaard, U. Magnea and S. Nishigaki, 
	{\em Nucl. Phys.} {\bf B487}, 721 (1997).

\bibitem{VERACTA}
J.J.M. Verbaarschot, 
	{\em Acta Phys. Polon.} {\bf B25}, 133 (1994).

\bibitem{TILO1}
T. Wettig, T. G\"{u}hr, A. Sch\"{a}fer and H.A. Weidenm\"{u}ller, 
	Hirschegg Lectures (1997), e-print hep-ph/9701387;
M.E. Berbenni-Bitsch et al., 
	e-print hep-lat/9704018.
 
\bibitem{PISARSKIW}
R. Pisarski and F. Wilczek, 
	{\em Phys. Rev.} {\bf D29}, 338 (1984). 

\bibitem{TILO2}
A.D. Jackson, M.K. Sener and J.J.M. Verbaarschot, 
	{\em Nucl. Phys.} {\bf B479}, 707 (1996). 

\bibitem{JNZ} 
J. Jurkiewicz, M. A. Nowak and I. Zahed,
        {\em Nucl. Phys.} {\bf B478}, 605 (1996); Erratum -{\em ibid.} 
{\bf B513}, 759 (1998).

\bibitem{PAUL2}
P.H. Damgaard and S.M. Nishigaki, 
	{\em Nucl. Phys.} {\bf B518}, (1998) 495;
%	e-print hep-th/9711023; 
T. Wilke, T. Guhr and T. Wettig, 
	{\em Phys. Rev.} {\bf D57}, (1998) 6486.
%	e-print hep-th/9711057.

\bibitem{WIGNER}
E.P. Wigner, 
	{\em Proc. Camb. Phi. Soc.} {\bf 47}, 790 (1951).   

\bibitem{JV} 
A.D. Jackson and J.J.M Verbaarschot,
	{\em Phys. Rev.} {\bf D53}, 7223 (1996).

\bibitem{PASTUR}
L.A. Pastur, 
	{\em Theor. Mat. Phys. (USSR)}  {\bf 10}, 67 (1972);
F. Wegner, 
	{\em Phys. Rev.} {\bf B19}, 783 (1979). 

\bibitem{STEPHANOV}
M. Stephanov, 
	{\em Phys. Rev. Lett.}  {\bf 76}, 4472 (1996);
M. Stephanov,
	{\em Nucl. Phys. Proc. Suppl.}  {\bf 53}, 469 (1997).

\bibitem{USMUX}
R.A. Janik, M.A. Nowak, G. Papp and I. Zahed, 
	{\em Phys. Rev. Lett.} {\bf 77}, 4876 (1996).

\bibitem{HUANG}
for a review, see e.g. K. Huang, 
	{\it Statistical Mechanics}, John Wiley, New York (1987).

\bibitem{HALASZ}
A. Halasz, A.D. Jackson and J.J.M. Verbaarschot,
	{\em Phys. Lett.} {\bf B395}, 293 (1997); 
A. Halasz, A.D. Jackson and J.J.M. Verbaarschot,
	{\em Phys. Rev.} {\bf D56}, 5140 (1997).

\bibitem{WEISE} 
U. Vogel and W. Weise,
	{\em Prog. Part. and Nucl. Phys.} {\bf 27}, 195 (1991).
 
\bibitem{ZHUANG} 
P. Zhuang, J. H\"ufner and S.P. Klevansky,
        Nucl. Phys. {\bf A576}, (1994) 525.

\bibitem{ALKOF}
R. Alkofer, M.A. Nowak, J.J.M. Verbaarschot and I. Zahed, 
	{\em Phys. Lett.} { \bf B233}, 205 (1989).  

\bibitem{WEISS}
D. Diakonov, M.V. Polyakov and C. Weiss, 
	{\em Nucl. Phys.}{\bf B461}, 539 (1996).

\bibitem{USU1}
R.A. Janik, M.A. Nowak, G. Papp and I. Zahed, 
	{\em Nucl. Phys.} {\bf B498}, 313 (1997).
 

\bibitem{USTHETA} 
R.A. Janik, M.A. Nowak, G. Papp and I. Zahed,
	to be published.

\bibitem{WITTEN}
E. Witten, 
	{\em Ann. Phys.} {\bf 128}, 363 (1980).
 
\bibitem{SCHIERHOLZ}
G. Schierholz,
        Nucl. Phys. Proc. Suppl. {\bf A37}, (1994) 203.

\bibitem{REVIEWMAT}
For an excellent  recent review, see 
T. Guhr, A. M\"{u}ller-Gr\"{o}ling and H.A. Weidenm\"{u}ller,
	{\em Phys. Rept.} {\bf 299}, (1998) 189.
%	{\em  e-print} cond-mat/9707301.

\bibitem{USDIAG}
R.A. Janik, M.A. Nowak, G. Papp and I. Zahed,
	{\em Nucl. Phys.} {\bf B501}, 603 (1997).

\bibitem{ZEEFE}
J. Feinberg and A. Zee, 
	{\em Nucl. Phys.} {\bf B501}, 643 (1997).

\bibitem{SHADES}
R.A. Janik, M.A. Nowak, G. Papp and I. Zahed, 
	{\em Acta Phys. Polon.} {\bf B28}, 2949 (1997).

\bibitem{JANIKPHD}
R.A. Janik, 
	Ph.D. Thesis, Cracow 1997 (unpublished). 

\bibitem{VINK} 
J. Vink, 
	{\em Nucl. Phys.} {\bf B323}, 399 (1989) 

\bibitem{USMAP}
R.A. Janik, M.A. Nowak, G. Papp, J. Wambach and I. Zahed, 
	{\em Phys. Rev.} {\bf E55}, 4100 (1997).
\end{thebibliography}
\end{document}